\pdfoutput=1 
\documentclass[aps,twocolumn,prl,amsmath,amssymb,10pt]{revtex4-1}
\usepackage{graphicx}
\usepackage{scrextend}
\usepackage{manfnt,pifont}
\usepackage{hyperref}
\usepackage{url}
\usepackage{xcolor}
\usepackage{url}
\usepackage{cleveref}
\usepackage[normalem]{ulem}
\usepackage{tikz}

\usetikzlibrary{decorations.markings}
\usetikzlibrary{arrows.meta}
\usetikzlibrary{calc}
\usetikzlibrary{bending}
\usetikzlibrary{snakes}

\begin{document}
%%%%%%%%%%%%%%%%%%%%%%%%%%%%%%%%%%%%%%%%%%%%%

\newcommand{\wb}[1]{  \textbf{\textcolor{red}{(#1 --wb)}}}
\newcommand{\kostas}[1]{  \textbf{\textcolor{blue}{(#1 --kostas)}}}
\newcommand{\ellis}[1]{  \textbf{\textcolor{purple}{(#1 --Ellis)}}}
\newcommand{\revision}[1]{\marginpar{\scriptsize\color{red}{\bfseries [Revision Note]}\\ #1}}

\title{Coon unitarity via partial waves or: how I learned to stop worrying and love the harmonic numbers}
%%%%%%%%%%%%%%%%%%%%%%%%%%%%%%%%%%%%%%%%%%%%%
\author{Konstantinos C. Rigatos}
\email{rkc@ucas.ac.cn}
\affiliation{
Kavli Institute for Theoretical Sciences, University of Chinese Academy of Sciences, Beijing 100190, China
}
\author{Bo Wang}
\email{b\_w@zju.edu.cn}
\affiliation{
Zhejiang Institute of Modern Physics, School of Physics, Zhejiang University, Hangzhou, Zhejiang 310058, China
\\
}
\affiliation{
Joint Center for Quanta-to-Cosmos Physics, Zhejiang University, Hangzhou, Zhejiang 310058, China
}
%%%%%%%%%%%%%%%%%%%%%%%%%%%%%%%%%%%%%%%%%%%%%
\date{\today}
%%%%%%%%%%%%%%%%%%%%%%%%%%%%%%%%%%%%%%%%%%%%%
\begin{abstract}
We present a novel approach to partial-wave unitarity that bypasses a lot of technical difficulties of previous approaches. In passing, we explicitly demonstrate that our approach provides a very suggestive form for the partial-wave coefficients in a natural way. We use the Coon amplitudes to exemplify this method and show how it allows to make important properties such as partial-wave unitarity manifest.
\end{abstract}
%%%%%%%%%%%%%%%%%%%%%%%%%%%%%%%%%%%%%%%%%%%%%
\maketitle
%%%%%%%%%%%%%%%%%%%%%%%%%%%%%%%%%%%%%%%%%%%%%
\noindent{\bf Introduction.} String amplitudes exhibit fascinating properties; tame high-energy behaviour, dual resonance and an infinite spin tower. The epitome of this triplet is the Veneziano amplitude \cite{Veneziano:1968yb}. Questions on its uniqueness were raised shortly after its discovery, at which time Coon suggested a deformation \cite{Coon:1969yw,Coon:1972qz,Baker:1976en} exhibiting all of the aforementioned properties, but with logarithmic Regge trajectories instead of linear ones, interpolating between the Veneziano and scalar theory amplitudes.

Recently it was put forth as an interesting case, since it evades the universal behaviour of linear Regge trajectories \cite{Caron-Huot:2016icg}. Several of its aspects are by now well-understood \cite{Figueroa:2022onw,Chakravarty:2022vrp,Bhardwaj:2022lbz,Geiser:2022icl,Jepsen:2023sia,Li:2023hce,Bhardwaj:2023eus,Geiser:2023qqq}, including rigorous unitarity bounds on the $\{q, m^2,D\}$-space with $q$ the deformation parameter of the Coon amplitude, $m^2$ the mass of external particles, and $D$ the number of dimensions. Further, there are qualitative similarities with open-string scattering on D-branes in AdS \cite{Maldacena:2022ckr}. 

With two explicit examples exhibiting stringy behaviour, a reasonable follow-up was to examine if there are more amplitudes with stringy characteristics and more specifically what are the consequences of imposing an appropriately tuned spectrum, additionally to the hallmark string properties. Naively, by demanding stringy properties and the string theory spectrum, we would expect that the answer would be uniquely fixed and quite surprisingly recent studies proved that this is not the case \cite{Cheung:2023adk,Cheung:2022mkw,Geiser:2022exp}. 

These new amplitudes were bootstrapped in a bottom-up way, and hence, we lack the underlying physics theories, if any. Thus, we have to turn to further consistency conditions to derive more robust statements and this is where unitarity enters. It is plausible that some S-matrix bootstrap solutions are just mathematical answers inconsistent with unitarity. 

Partial-wave unitarity is highly non-trivial, even for the simple Veneziano amplitude \cite{Maity:2021obe,Arkani-Hamed:2022gsa}. We still miss a satisfactory derivation of the critical dimension of the (super)string directly from an analysis of tree-level scattering, see \cite{Arkani-Hamed:2022gsa} for recent work and progress. It becomes an even more pressing matter in cases without an underlying theory, as it imposes non-trivial bounds on the free parameters of the amplitudes.

We provide a different approach to partial-wave unitarity\footnote{See also the recent \cite{vanRees:2023fcf,Eckner:2024ggx} for interesting approaches based on analytics and numerics.}. The keypoint of our analysis resides in packaging crucial information into harmonic numbers and reducing the analysis to a linear algebra problem. We choose the Coon amplitude, to demonstrate that our method works well even when dealing with non-trivial deformations, such as the $q$-deformation. 
~\\

\noindent{\bf The Coon amplitude.} The Coon amplitude is given by \cite{Coon:1969yw} 
\begin{equation}\label{eq:coon_def}
\mathcal{A}(s,t) = (q-1) q^{\tfrac{\log \sigma}{\log q} \tfrac{\log \tau}{\log q}} \prod^{\infty}_{n=0} \frac{(\sigma \tau - q^{n})(1-q^{n+1})}{(\sigma-q^n)(\tau-q^n)}
\,			,
\end{equation}
with the deformation parameter taking values in the range $q \in (0,1)$ and we have used the abbreviations
\begin{equation}\label{st_defs}
\sigma = 1 + (s-m^2)(q-1)
\,		,
\quad
\tau = 1 + (t-m^2)(q-1)
\,		.
\end{equation}
The $s$-channel poles are located at $\sigma = q^n$, or equivalently 
\begin{equation}\label{eq:schannel_poles}
s_N = m^2 + [N]_q
\,		,
\end{equation}
where in the above $[N]_q$ is the $q$-deformed integer. Our definition is:
\begin{equation}\label{eq: q_int_def}
[\mathfrak{a}]_{\mathfrak{b}} = \frac{1-\mathfrak{b}^{\mathfrak{a}}}{1-\mathfrak{b}}
\,      .
\end{equation}
Note that we are using the same conventions as \cite{Figueroa:2022onw,Chakravarty:2022vrp} and hence in comparing against the results of \cite{Bhardwaj:2022lbz} one has to be cautious with some shifts. 

The $N \rightarrow \infty$ limit reveals an accumulation point:
\begin{equation}\label{eq:accumulationpoint}
s_{*} = m^2 + \frac{1}{1-q}
\,		.
\end{equation}
Denoting by $\mathcal{R}(t,q,N)$ the residue of \cref{eq:coon_def} at the s-channel poles we have
\begin{equation}\label{eq:res_final}
\mathcal{R}(t,q,N)
= \frac{q^N}{\left(q^{-N};q\right){}_{N}} ~ \sum^{N}_{n=0} \frac{\left(q^{-N};q\right)_n}{(q;q)_n} ~ \tau^{N-n}
\,			,
\end{equation}
where $(x;q)_{n}$ is the $q$-Pochhammer symbol defined via \cref{eq: qpoch_def}:
\begin{equation}\label{eq: qpoch_def}
    (x;q)_n\equiv \prod_{i=0}^n(1-x\; q^i)\, .
\end{equation}

We proceed by expanding \cref{eq:res_final} in a basis of Gegenbauer polynomials:
\begin{equation}\label{eq:res_gegen}
\mathcal{R}(t,q,N) =  \sum_{\ell=0}^{N} c_{N,\ell} 
C_{\ell}^{(\alpha)} \left( 1+ \tfrac{2t}{s_N-4m^2}\right)
\,			,
\end{equation}
where $\alpha=\tfrac{D-3}{2}$, $D$ is the spacetime dimensions, and we will use the shorthand $\mathcal{N} =s_N-4m^2= [N]_q - 3m^2$.

The necessary and sufficient condition for unitarity is
\begin{equation}
c_{N,\ell} \geq 0
\,          ,
\end{equation}
such that the spectrum is free of negative-norm states.

To proceed, we exploit the fact that the Gegenbauer polynomials form a complete and orthogonal basis
\begin{equation}\label{eq:gegen_ortho}
\int_{-1}^{+1} dx C_{\ell}^{(\alpha)} (x) C_{\ell^{\prime}}^{(\alpha)}(x)
(1-x^2)^{\alpha-\frac{1}{2}} = 2 \mathcal{K}(\ell,\alpha) \delta_{\ell \ell^{\prime}}
\,			,
\end{equation}
where $\mathcal{K}(\ell,\alpha)$ is the normalization factor
\begin{equation}\label{eq:normalizationgegen}
\mathcal{K}(\ell,\alpha) = \frac{\pi\Gamma(\ell+2\alpha)}{2^{2\alpha}\ell! (\ell+\alpha) \Gamma^2(\alpha)}
\,      .
\end{equation}
This allows us to obtain an integral representation formula of the partial-wave coefficients; note that we shift $n\to N-r$ in \cref{eq:res_final}:
\begin{equation}\label{eq:pwintegral}
\begin{aligned}
&c_{N,\ell} =
\frac{4^{\alpha-\tfrac{1}{2}}}{\mathcal{N}^{2\alpha}}
\frac{1}{\mathcal{K}(\ell,\alpha)}
\frac{q^N}{\left(q^{-N};q\right){}_{N}} 
\sum^{N}_{r=0} \frac{\left(q^{-N};q\right)_{N-r}}{(q;q)_{N-r}}
\\
&\int\limits^{0}_{-\mathcal{N}}dt C_{\ell}^{(\alpha)} \left( 1+ \tfrac{2t}{\mathcal{N}}\right)
[(1-q)(m^2-t)+1]^r [t(-t-\mathcal{N})]^{\alpha-\tfrac{1}{2}}
\,           .
\end{aligned}
\end{equation} 
~\\

\noindent{\bf Generating function for the integral.} We are going to use a well-established approach to derive the expression for the partial-wave coefficients. We briefly describe the basic steps and the interested reader can find thorough explanations of this method in  \cite{Chakravarty:2022vrp,Maity:2021obe,Rigatos:2023asb}. We proceed, by making use of the generating function of the Gegenbauer polynomials \cref{eq: gen_function_gegenbauer} and the binomial expansion in order to express the $[t(-t-\mathcal{N})]^{\alpha-\tfrac{1}{2}}$ term as powers of $t$. This results in the following expression:
\begin{equation}\label{eq: main_eq_aNl}
    \begin{aligned} 
&\sum^{\infty}_{j=0} \mathcal{K}(j,\alpha) c_{N,j}h^j = 
\frac{4^{\alpha-\tfrac{1}{2}}}{\mathcal{N}^{2\alpha}}
\frac{q^N}{\left(q^{-N};q\right){}_{N}} 
\\
&\sum^{N}_{r=0} \frac{\left(q^{-N};q\right)_{N-r}}{(q;q)_{N-r}}
\sum^{\alpha-\tfrac{1}{2}}_{\mathfrak{p}=0} \binom{\alpha-\tfrac{1}{2}}{\mathfrak{p}} (-\mathcal{N})^{\alpha-\mathfrak{p}-\tfrac{1}{2}} ~ \mathcal{G}
\,      ,
    \end{aligned}
\end{equation}
where in the above we have defined a ``pseudo-generating function" to be given by:
\begin{equation}\label{eq: gen_function_def}
\mathcal{G} = \int\limits^{0}_{-\mathcal{N}} dt \frac{\left(-(1-q)t+1+m^2(1-q)\right)^r t^{\alpha+\mathfrak{p}+\tfrac{1}{2}}}{\left((h-1)^2 - \frac{4ht}{\mathcal{N}}\right)^{\alpha}}
\,          .
\end{equation}
The integral in \cref{eq: gen_function_def} can be performed analytically in terms of the Appel hypergeometrics function and yields:
\begin{equation}
    \begin{aligned}
&\mathcal{G} = 
- 
\frac{2(-\mathcal{N})^{\alpha+\mathfrak{p}+\tfrac{1}{2}}}{1+2\alpha+2\mathfrak{p}}
\frac{(1+m^2(1-q))}{(h-1)^{2\alpha}}
\\
&F_{1}
\left(\alpha+\mathfrak{p}+\tfrac{1}{2};-r,\alpha;\alpha+\mathfrak{p}+\tfrac{3}{2}; -\frac{\mathcal{N}}{s_{*}}, -\frac{4h}{(h-1)^2} \right)
\,          .
    \end{aligned}
\end{equation}
We proceed further by using the definition of the Appell function as a formal power-series \cref{eq: def_F1_main} and the binomial expansion for the $h$-dependent term in order to derive: 
\begin{equation}\label{eq: final_cal_G}
    \begin{aligned}
&\mathcal{G} =
-2^{r+1} (-\mathcal{N})^{\alpha+\mathfrak{p}+\tfrac{1}{2}}(1+m^2(1-q))^r 
\\
&\sum^{r}_{\mathfrak{w}=0} \frac{(\alpha)_{\mathfrak{w}}}{\mathfrak{w_2}!} \frac{(-4)^{\mathfrak{w}}(-1)^{-2\mathfrak{w}-2\alpha}}{1+2\alpha+2p+2\mathfrak{w}} 
\\
&{}_2F_{1} \left(-r,\alpha + \mathfrak{p} + \mathfrak{w} + \frac{1}{2}, \alpha + \mathfrak{p} + \mathfrak{w} + \frac{3}{2}; -\frac{\mathcal{N}}{s_{*}} \right)
\\
&h^{\mathfrak{w}}
\sum^{\infty}_{\mathfrak{e}} \frac{(2 \mathfrak{w} + 2 \alpha)_{\mathfrak{e}}}{\mathfrak{e}!} h^{\mathfrak{e}}
\,              .
    \end{aligned}
\end{equation}
Substituting \cref{eq: final_cal_G} into \cref{eq: main_eq_aNl} we have expressed both sides as powers of $h$ and we simply wish to extract the $h^{\ell}$ term from both sides that gives the $c_{N,\ell}$ coefficients. The final result is: 
\begin{equation}\label{eq: cNl_gen_fun}
    \begin{aligned}
&c_{N,\ell}=
\frac{4^{\alpha}}{\mathcal{K}(\ell,\alpha)} 
\frac{q^N}{\left(q^{-N};q\right){}_{N}} 
\\
&\sum^{N}_{r=0}\sum^{\alpha-\tfrac{1}{2}}_{\mathfrak{p}=0}\sum^{\ell}_{\mathfrak{w}=0} 
\frac{\left(q^{-N};q\right)_{N-r}}{(q;q)_{N-r}} 
(1 + m^2(1-q))^r 
\\
&\binom{\alpha-\tfrac{1}{2}}{\mathfrak{p}}(-1)^{\mathfrak{p}} 
\frac{(-4)^{\mathfrak{w}}(\alpha)_{\mathfrak{w}} (2\mathfrak{w}+2\alpha)_{\ell-\mathfrak{w}}}{\mathfrak{w}!(1+2\alpha+2p+2\mathfrak{w})(\ell-\mathfrak{w})!} 
\\
&{}_2F_{1}
\left(
-r,\alpha+\mathfrak{p}+\mathfrak{2}+\tfrac{1}{2}, \alpha+\mathfrak{p}+\mathfrak{2}+\tfrac{3}{2}; -\frac{\mathcal{N}}{s_{*}}
\right)
\,          .
    \end{aligned}
\end{equation}
We have, thus, derived a representation of all partial-wave coefficients in terms of triple truncated sums. While it might appear a bit more cumbersome than the original integral representation, \cref{eq:pwintegral}, it is algorithmically simpler to consider. However, we proceed to demonstrate an even simpler representation of the partial-wave coefficients.  
~\\

\noindent{\bf Discussing other approaches.} We would like to comment on \cref{eq: cNl_gen_fun} and related results in the literature. In \cite[eq.(67)]{Cheung:2022mkw} the authors presented a formula for all partial-wave coefficients of the Coon amplitude as a nested fourfold sum. Compared to that way of writing the partial-wave coefficients \cref{eq: cNl_gen_fun} appears to be simpler. Also, in \cite{Bhardwaj:2022lbz} the authors used the contour integral representation, that was originally developed in \cite{Arkani-Hamed:2022gsa} for the Veneziano amplitude, to solve for the partial-wave coefficients. However, an explicit expression for all spins and mass levels was not derived in closed-form. The authors in \cite{Chakravarty:2022vrp} derived the expressions for the leading Regge trajectory and the general partial-wave coefficients but only for the $D=4$ case. Hence, \cref{eq: cNl_gen_fun} presents itself as the simpler and most general result so far. As we shall see, we will be able to derive a much simpler expression even compared to \cref{eq: cNl_gen_fun} below.
~\\

\noindent{\bf Intermezzo: harmonic numbers.} Before we proceed, we give some very basic facts about harmonic numbers. This is the set that will serve as basis in the analysis below. 

We define the $q$-deform harmonic number
\begin{equation}\label{eq: hm}
    Z^q_{\{i_1,i_2,\cdots,i_k\}}(N) \equiv \sum_{n=1}^{N} \frac{q ^{n} }{[n]_q^{i_1}}Z^q_{\{i_2,i_3,\cdots,i_k\}}(n-1)\,  ,
\end{equation}
where $i$ is the symbol. We also define $Z^q_{\{\empty\}}(N)=1$ and $Z^{q}_{\{i_1,i_2,\cdots,i_k\}}(0)=0$ with $k\geq1$. These are also called multiple harmonic $q$-series\cite{Bradley_2005}.

In our context, we only encounter the condition that every symbol $i$ is $1$. Therefore, we introduce the notation
\begin{equation}\label{eq:notationZq}
    Z_{k}^{q}(N)=Z^q_{\{1,1,\cdots,1\}}(N) \,  , 
\end{equation}
where the length of symbol list is $k$. More useful properties of $q$-deform harmonic numbers can be found in Supplemental Material.
~\\

\noindent{\bf Sum rules and harmonic numbers.} Next we provide a different approach to determine all partial-wave coefficients. We will bypass many difficulties in the generating function method. This method also offers a novel perspective that allows us to encode information into harmonic numbers and then simply obtain all the coefficients by solving some linear equations.

First of all, we need to re-arrange  the form of the residues slightly. Following \cite{Geiser:2022icl}, we re-organize \cref{eq:res_final} as
\begin{equation}\label{eq:reshm}
     \mathcal{R}(t,q,N)= {q^N} \prod_{n=1}^{N} \left(1+(t-m^2) \frac{q^n}{[n]_q}\right).
\end{equation}
The next observation is crucial. Let us consider the series coefficient of $t$, which leads us to the key identity
\begin{equation}
     \prod_{n=1}^{N} \left(1+t \frac{q^n}{[n]_q}\right) = \sum_{n=0}^{N} t^{n} Z^q_{ {n}}(N)
\, ,
\end{equation}
where $Z^q_{ {n}}(N)$ is the $q$-deformed multiple harmonic number; for the definition see \cref{eq: hm} and \cref{eq:notationZq} in Supplemental Material. In fact, we treat the residue $\mathcal{R}(t,q,N)$ as a generating function of $t$ with coefficients $Z^q_{ {n}}(N)$. 

By using the binomial theorem, we straightforwardly derive the series expansion of $\mathcal{R}(t,q,N)$ as a series in $t$
\begin{equation}
     \mathcal{R}(t,q,N)= {q^N} \sum_{n,k=0}^{N} \binom{k}{n} (-m^2)^{k-n} t^{n} Z^q_{ {k}}(N)\, .
\end{equation}

On the other hand, let us revisit the decomposition of the Gegenbauer polynomials. By substituting
\begin{equation}\label{eq: ident_gegen}
    C^{(\alpha)}_\ell(x)=
\frac{(2\alpha)_{\ell}~\;}{\Gamma(\ell+1)} 
{}_2F_1\left(-\ell,\ell+2\alpha;\alpha+\frac{1}{2};\frac{1-x}{2}\right)
\,     
.
\end{equation}
into \cref{eq:res_gegen} we get:
\begin{align}\label{eq: reswithT}
    \mathcal{R}(t,q,N) &=\sum_{n,\ell=0}^{N} c_{N,\ell} \mathcal{T}_{n,\ell} \left(-\frac{t}{\mathcal{N}}\right )^n\, ,
\end{align}
where
\begin{equation}
     \mathcal{T}_{n,\ell}=\frac{\Gamma (\ell+2\alpha) (-\ell)_n (\ell+2\alpha)_n}{n!\, \ell! \, \Gamma (2\alpha) \left(\frac{1}{2}+\alpha\right)_n}\, .
\end{equation}

So far we have obtained two expansions of $\mathcal{R}(t,q,N)$: one is in the basis of harmonic numbers and the other is the decomposition of the partial-wave coefficients. We can identity them, and since they are both series in $t$ we arrive at the sum rules equations
\begin{equation}
    \sum_{\ell=0}^{N} c_{N,\ell} \mathcal{T}_{n,\ell} \left(-\frac{1}{\mathcal{N}}\right )^n ={q^N} \sum_{k=0}^{N} \binom{k}{n} (-m^2)^{k-n}  Z^q_{ {k}}(N)\, .
\end{equation}
Ergo the problem has reduced to an elementary linear algebra exercise the solution of which is given by:
\begin{align}\label{eq:res_doublehm}
     c_{N,\ell} ={q^N}& \sum_{n,k=0}^{N} \underbrace{\binom{n}{\ell} \frac{\sqrt{\pi} }{\mathcal{K}(\ell,\alpha)} 
     \frac{(-1)^{\ell} (\alpha)_{\frac{1}{2}+n}}{(\ell+2\alpha)_{1+n}}  }_{\mathcal{T}_{n,\ell}^{-1}} \nonumber\\
     &\underbrace{\binom{k}{n} (-m^2)^{k-n}}_{\text{external mass}} \underbrace{ {\left(3 m^2-[N]_q\right)^n} \vphantom{\binom{a}{n}}}_{\text{scattering angle}} Z^q_{ {k}}(N)\, .
\end{align}
The structure of \cref{eq:res_doublehm} suggests a natural explanation of each term. The first part, $\mathcal{T}_{n,\ell}^{-1}$, is just the inverse of the matrix $\mathcal{T}_{n,\ell}$ and is universal. The second is controlled by the external mass and reduces to 1 for massless particles. The third part is related to scattering angle and the last part is our $q$-deformed harmonic number serving as a basis. 

We point out that the summation over $n$ in \cref{eq:res_doublehm} can be performed analytically. Since the binomial coefficient becomes $0$ when $n>k$, we can extend the upper bound of $n$ to infinity. Thus, we finally obtain a single-sum version
\begin{align}\label{eq:res_singlehm}
    c&  _{N,\ell}=q^{N} \sum_{k=0}^{N}2 (-1)^{\ell } \Gamma (2 \alpha ) (-m^2)^k  (\alpha +\ell ) Z_{ {k}}^q(N) \nonumber \\
    & \, _3\tilde{F}_2\left(1,\alpha +\frac{1}{2},-k;1-\ell ,\ell +2 \alpha +1;-\frac{\mathcal{N}}{m^2}\right)
\,      ,
\end{align}
where we have used ${}_{p}\tilde{F}_{q}(a_1,\ldots,a_p;b_1,\ldots,b_q;z)$ to denote the regularised hypergeometric function, for the definition see \cref{eq:defreghyp}. 
~\\

\noindent{\bf Analysis.} Having pursued a new way to compute the partial-wave coefficients, we make some consistency checks. Firstly, it is easy to check that the expressions in \cref{eq:pwintegral,eq: cNl_gen_fun,eq:res_doublehm} agree with one another. Further, it is straightforward to check that for the special case $D=4$ there is agreement with the leading Regge trajectory and the general formula derived in \cite{Chakravarty:2022vrp}. We have, also, checked our results against the coefficients \cite[eq.(42)]{Figueroa:2022onw}; these are explicit checks up to level 3. 

The next comment pertains to the structure of \cref{eq:res_doublehm}. Notice how much more suggestive the answer is compared to the integral representation, \cref{eq:pwintegral}, and the hypergeometric sum expression, \cref{eq: cNl_gen_fun}.

\begin{figure}[t]
    \centering
    \includegraphics[scale=0.6]{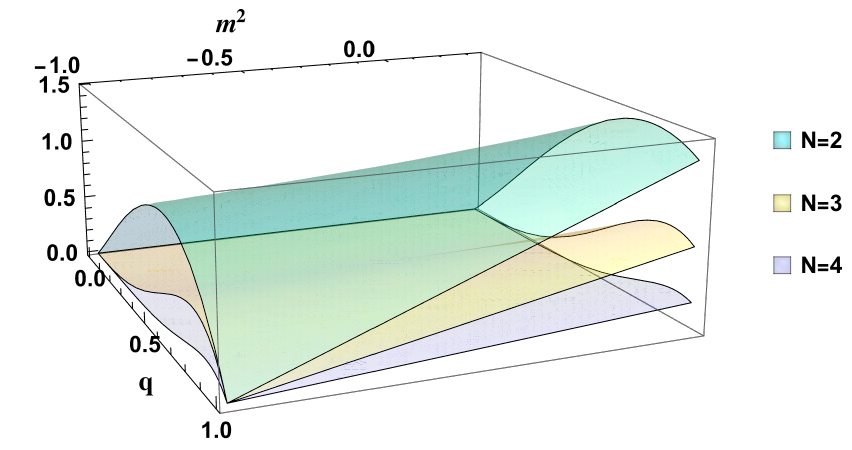}
    \caption{The sub-leading trajectory is controlled by function $f(q,m^2)$. Let us consider $0<q<1$ and $-1\leq m^2\leq \frac{1}{3}$. We find that the roots of $f(q,m^2)=0$ are at $q=0$ or $q=1$. We choose $N=2,3,4$ as a vivid illustration.}
    \label{fig:subleading}
\end{figure}

We proceed to examine some unitarity bounds. 

Using the basis of harmonic numbers facilitates the analysis of Regge trajectories. We present the formulae for the leading and sub-leading trajectories to illustrate. For the former we have the simple
\begin{equation}
    c_{N,N}=\left(\frac{q}{4}\right)^{N}\frac{ N! }{(\alpha )_{N}} \mathcal{N}^{N} Z^q_{ {N}}(N)\, .
\end{equation}
Requiring unitarity yields
\begin{equation}
    \mathcal{N}=[N]_q-3m^2\geq0,\quad  \forall\; N\geq 1\, .
\end{equation}
This is our first non-trivial bound. When $N=1$, we easily obtain $m^2\leq \frac{1}{3}$. It is well known that the Veneziano amplitude is non-unitary when $m^2<-1$. Taking this into consideration, we set $-1\leq m^2\leq \frac{1}{3}$ as our parameter interval. 

Next let us focus on the sub-leading trajectory. From \cref{eq:res_doublehm} we have
\begin{align}\label{eq:subleading}
    c_{N,N-1} \propto \underbrace{2 Z^q_{ {N-1}}(N)-N(2 m^2+\mathcal{N})  Z^q_{ {N}}(N)}_{f({q,m^2})}\, .
\end{align}
Ignoring an overall positive factor, and using the explicit form of harmonic numbers, \cref{eq:Zexplicitform}, we find that the locus of zeroes is on the boundary of the $q$-region; see \cref{fig:subleading}. By taking the special limit $q\to 1$, we notice
\begin{equation}
    f(1,m^2)=\frac{m^2+1}{\Gamma (N)}\geq 0\, .
\end{equation}
Hence we prove that the sub-leading trajectory is \textit{always manifestly positive}.

The use of harmonic numbers, also, provides a significant benefit in low-spin analysis. We choose $D=4$ and $\ell=0$ as an example to support this claim. The partial-wave coefficient \cref{eq:res_singlehm} becomes
\begin{align}\label{eq: corrected}
    c_{N,0}&=\frac{1}{\mathcal{N}} \frac{q^N}{\left(q^{-N};q\right){}_{N} } \sum_{j=0}^N  \frac{\left(q^{-N};q\right){}_{N-j}}{(q;q)_{N-j}} \frac{1}{(1-q)(1+j) }  \nonumber \\
            &\left[\left((m^2+\mathcal{N}) (1-q)+1\right)^{j+1}-\left(m^2 (1-q)+1\right)^{j+1}\right]\, .
\end{align}
The first line not always positive, however, we know \cite{Wang:2024wcc} that the low-spin states dominate in the unitarity analysis. The second-line of \cref{eq: corrected} is \textit{manifestly positive} due to $(m^2+\mathcal{N}) (1-q)+1\geq m^2 (1-q)+1\geq0$. Using this and some mild numerical analysis, we observe that the partial-wave coefficient $c_{N,0}$ is positive when $D=4$.  

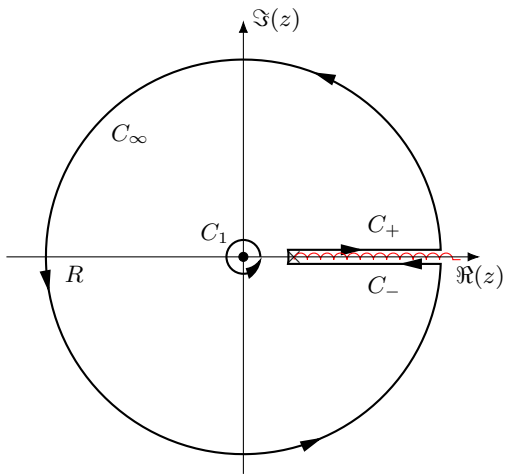
\begin{figure}[t]
    \centering
    \begin{tikzpicture}[>={Kite[inset=0pt,length=0.3cm,bend]},
      decoration={markings,
      mark= at position 0.15 with {\arrow{>}},
      mark= at position 0.42 with {\arrow{>}},
      mark= at position 0.65 with {\arrow{>}},
      mark= at position 0.82 with {\arrow{>}},
      mark= at position 0.95 with {\arrow{>}},
      },scale=0.75]

% Configurable parameters
\def\gap{0.25}
\def\bigradius{3.5}
\def\littleradius{0.8}
\def\lradius{0.3}

% Path1
\draw[black, thick,   decoration, 
      postaction={decorate}]  
  let
     \n1 = {asin(\gap/2/\bigradius)},
     \n2 = {asin(\gap/2/\littleradius)}
  in (\n1:\bigradius) arc (\n1:360-\n1:\bigradius) 
  -- (-\n2:\littleradius) arc (-\n2:0+\n2:\littleradius)
  -- cycle;

% Path2
\draw[black, thick, ->] 
      (\lradius,0) arc (0:360:\lradius);

%Label
\node[above left] at (-0.1,0.1) {$C_1$};
\node[below right] at (-2.5,2.5) {$C_{\infty}$};
\node[above] at (2.5,0.2) {$C_+$};
\node[below] at (2.5,-0.2) {$C_-$};
\node[below] at (-\bigradius+0.5,0) {$R$};

%Branch
\node[] at (\littleradius+0.1,0) {$\times$};
\draw[snake=bumps,color=red] (\littleradius+0.1,-0.05) -- (1.1*\bigradius,-0.05);

\draw[fill=black] (0,0) circle(0.08);

% axes
\draw[-Latex] (-1.2*\bigradius,0) -- (1.2*\bigradius,0) node[below]{$\Re(z)$} ;
\draw[-Latex] (0,-1.1*\bigradius) -- (0,1.2*\bigradius) node[right]{$\Im(z)$};

\end{tikzpicture}
    \caption{The complex plane. The original contour is $C_1$ and the deformed contour is $C_\infty+C_-+C_+$. We use red bumps to label the  branch cut.}
    \label{fig:contour}
\end{figure}

In some special limits, we can analyse the \textit{full} unitarity behavior in a new integral representation. We consider the limit $q\to 1$ and $m^2=0$, which actually is the super-string case. Combine \cref{eq:res_doublehm} and \cref{H1ktoZ}, the partial-wave coefficients are
\begin{align}\label{eq:superstringint}
    c&_{N-1,\ell} =\frac{1}{2\pi i} \oint \frac{2(-1)^{\ell}}{z^{N}(1-z)} (\ell+\alpha)\Gamma(2\alpha) \nonumber\\
    & _2\tilde{F}_2 \left(1,\alpha+\frac{1}{2};1-\ell, \ell+2\alpha+1;N\log(1-z) \right)\, ,
\end{align}
there is a shift in $N$ for convenience \footnote{This result matches in \cite[Table 1]{Arkani-Hamed:2022gsa} up to an overall factor of $N$. We shift $N\to N+1$ in the scattering angle of \cref{eq:res_doublehm} then take the limit $q\to1$ and $m^2\to0$.}. The integral is to extract the coefficient before $z^{N-1}$. The contour $C_1$ around $z=0$ can be deformed into $C_\infty+C_-+C_+$, see \cref{fig:contour}. The contribution from $C_\infty$ goes to $0$ when $R\to\infty$. There is also a branch cut from $\log(1-z)$, hence we convert \cref{eq:superstringint} into the integral of its Disc. For example, we choose $D=6$ and $\ell=0$ \footnote{It is well known that superstring partial-wave coefficient $c_{N-1,\ell}=0$ when $N-1+\ell$ is odd. So we focus on $N=1,3,5,\cdots$ here.}, 
\begin{align}\label{eq:superstringd6l0}
    c_{N-1,0}=\int_{0}^{\infty}dt \frac{12 \left(g(t)+\pi ^2 (t^{N}+1)\right)}{t\; N^3 (t+1)^{N} \left(\log ^2(t)+\pi ^2\right)^3}\, .
\end{align}
Here $t=z-1$ and 
\begin{equation}
    g(t)=\log (t) \left[N \left(t^{N}-1\right) \left(\log ^2(t)+\pi ^2\right)-3 \left(t^{N}+1\right) \log(t)\right].
\end{equation}
The integrand of \cref{eq:superstringd6l0} is always positive when $t\geq0$. The only subtlety is the function $g(t)$. However, notice that $g(t)$ has only one root, at $t=1$, and $g\rightarrow\infty$ when $t\to0$ and $t\to\infty$. We conclude that $g(t)\geq0$ in the region of integration. A more in-depth analysis with more examples is presented in \cite{Wang:2024wcc}.

Let us comment on the critical $q$. This is a value for $q$ below which the Coon amplitude is unitary for any choice of $D$ and $m^2$. Using the large-$N$ behaviour of the harmonic numbers we derive:
\begin{align}\label{eq: large_N}
    \lim_{N\to\infty }c_{N,0}&\propto \lim_{N\to\infty } \left[  q^N-\frac{N}{2}  q^N \left(-3 m^2 +[N]_q\right)  Z^{q}_1(N)  \right] \nonumber \\
    &\propto2 q^{-1}-m^2 (q-1)-3\; .
\end{align}
Setting \cref{eq: large_N} to zero, we conclude that the critical point for $q$ is:
\begin{equation}
    q_\infty(m^2)=\frac{m^2-3+\sqrt{9+2m^2+m^4}}{2m^2}\, ,
\end{equation}
which matches exactly $q_\infty(m^2)$ from \cite{Figueroa:2022onw}.
~\\

\noindent{\bf Discussion.} Motivated by the current status in the S-matrix bootstrap and the importance of partial-wave unitarity in these analyses we have suggested a novel method for obtaining an analytic expression for all partial-wave coefficients. 

One advantage of this method is the naturally structured and very suggestive form of the answer. Another is that it provides a way to solve for the coefficients by considering a linear algebra problem, much simpler than the original. A third upshot is the use of harmonic numbers and deformations thereof, which allows to analyse non-trivially deformed amplitudes in a straightforward manner. Furthermore, due to the properties of the harmonic numbers the simplified cases of the deformations follow by definition. Moreover, this new approach is very powerful in the low-spin trajectories. And finally, we can exploit the well-established asymptotic behaviour of harmonic numbers to obtain critical values for the various deformation parameters; here it was the $q$-deformation.

Our approach is easily generalisable to other cases, owing to its versatility. It is interesting to examine the implications and computational power on other amplitudes in order to carve out unitarity regions in the relevant parameter spaces. In \cite{Wang:2024wcc} one of us demonstrates the applicability of the method developed here in a more elaborate deformation, namely the hypergeometric Coon \cite{Cheung:2023adk}. As another future avenue it would be very interesting to understand if the basis of the harmonic numbers used in the partial-wave analysis, can be utilized in order to provide information about other observables or other setups, e.g., the Wilson coefficients \cite{Haring:2023zwu} or in studies in AdS backgrounds \cite{Alday:2022uxp,Alday:2022xwz}.
~\\

\noindent{\bf Acknowledgements.} We thank Zhongjie Huang and Ellis Ye Yuan for helpful comments on the draft. KCR is grateful to James Drummond for pointing out \cite{Figueroa:2022onw} some time ago and for related discussions, and also to Ellis Ye Yuan for for the warm hospitality and interesting discussions at Zhejiang University where parts of this work were performed. The work of KCR is supported by starting funds from University of Chinese Academy of Sciences (UCAS), the Kavli Institute for Theoretical Sciences (KITS), and the Fundamental Research Funds for the Central Universities. The work of BW is supported by National Science Foundation of China under Grant No.~12175197, Grant No.~11935013 and Grand No.~12347103. BW is also supported by the Fundamental Research Funds for the Chinese Central Universities under Grant No.~226-2022-00216.

\vspace{0.1cm}

\appendix
\section{{\large \sc{Supplemental Material}}}
Here we collect the basic formulae the were used in the main body of this work for the reader's convenience. 
    \subsection{Related to the generating function approach}\label{sec: gen_fun_app}
The generating function of the Gegenbauer polynomials is: 
\begin{equation}\label{eq: gen_function_gegenbauer}
\sum^{\infty}_{\ell=0} C^{(\alpha)}_{\ell}(x) t^{\ell} = \frac{1}{(1-2xt+t^2)^{\alpha}}			
\,			.
\end{equation}
The definition of the Appell hypergeometrics function as a formal power series is:
\begin{equation}\label{eq: def_F1_main}
F_{1}(\mathfrak{a};\mathfrak{b},\mathfrak{c};\mathfrak{d};x,y) = \sum^{\infty}_{\mathfrak{e}=0}\sum^{\infty}_{\mathfrak{f}=0}\frac{1}{\mathfrak{e}!}\frac{1}{\mathfrak{f}!} \frac{(\mathfrak{a})_{\mathfrak{e}+\mathfrak{f}}(\mathfrak{b})_{\mathfrak{e}}(\mathfrak{c})_{\mathfrak{f}}}{(\mathfrak{d})_{\mathfrak{e}+\mathfrak{f}}} x^{\mathfrak{e}} y^{\mathfrak{f}}			
\,			.    
\end{equation}

    \subsection{Related to the harmonic numbers method}\label{sec: harmonic_app}
Here we remind the reader the definition of the regularized hypergeometric function: 
\begin{equation}\label{eq:defreghyp}
\begin{aligned}
&{}_{p}\tilde{F}_{q}(a_1,a_2,\ldots,a_p;b_1,b_2,\ldots,b_q;z)
=
\\
&\qquad \frac{{}_{p}F_{q}(a_1,a_2,\ldots,a_p;b_1,b_2,\ldots,b_q;z)}{\Gamma(b_1)\Gamma(b_2)\ldots\Gamma(b_q)}
\,			.
\end{aligned}
\end{equation}
By equating \cref{eq:res_final} and \cref{eq:reshm} we obtain a simple expression for $Z^{q}_{ {k}}(N)$
\begin{align}
    Z_{ {k}}^{q}(N)=\frac{(q-1)^k }{\left(q^{-N};q\right){}_{N}} {\sum _{j=0}^{N}\binom{j}{k} \frac{ \left(q^{-N};q\right){}_{N-j}}{(q;q)_{N-j}}}\, .
\end{align}
This elegant identity plays a salient role in the unitarity analysis. Let us give some explicit examples,
\begin{align}\label{eq:Zexplicitform}
    Z^q_{N}(N)&= \frac{(q-1)^N }{\left(q^{-N};q\right){}_{N}}   \, , \\
    Z^q_{N-1}(N)&= \frac{(q-1)^N }{\left(q^{-N};q\right){}_{N}} \frac{ q^{-N} [N]_q -N}{1-q}.
\end{align}
The positivity of $Z^q_{k}(N)$ is not obvious in this form. However, we emphasize that the $q$-deformed harmonic number never be negative.

 In the limit $q\to1$, the $q$-deformed harmonic number returns to ordinary multiple harmonic number by definition, where $[n_j]_q\to n_j$. The ordinary multiple harmonic number is generating by harmonic polylogarithms\cite{Maitre:2005uu}, 
 \begin{equation}\label{H1ktoZ}
     \frac{1}{1-z}H(1_1,1_2,\cdots,1_k;z)=\sum_{N=0}^{\infty} Z^{1}_{k}(N) z^{N}\, ,
 \end{equation}
where dividing $1-z$ is for convenience. Actually, when every symbol is $1$ the polylogarithm $H$ is just a simply function
\begin{equation}\label{H1ktolog}
    H(1_1,1_2,\cdots,1_k;z)=\frac{(-1)^{k}}{k!}\log^k(1-z)\, .
\end{equation}
%%%%%%%%%%%%%%%%%%%%%%%%%%%%%%%%%%%%%%%%%%%%%
\bibliography{coonbib}
%%%%%%%%%%%%%%%%%%%%%%%%%%%%%%%%%%%%%%%%%%%%%
\end{document}